\documentstyle[aps,prl]{revtex}
\begin{document}
\twocolumn[\hsize\textwidth\columnwidth\hsize\csname @twocolumnfalse\endcsname

\title{Adsorption of benzene on Si(100) from first principles}
\author{Pier Luigi Silvestrelli, Francesco Ancilotto, and Flavio Toigo}
\address{Istituto Nazionale per la Fisica della Materia and
Dipartimento di Fisica ``G. Galilei'',
Universit\`a di Padova, via Marzolo 8, I-35131 Padova, Italy\\}

\date{\today}
\maketitle

\begin{abstract}
Adsorption of benzene on the Si(100) surface
is studied from first principles.
We find that the most stable configuration is a
tetra-$\sigma$-bonded structure characterized by 
one C-C double bond and four C-Si bonds.
A similar structure, obtained
by rotating the benzene molecule by 90$^{\circ}$, lies slightly higher 
in energy.
However, rather narrow wells on the potential energy surface 
characterize these adsorption configurations. 
A benzene molecule impinging on the 
Si surface is most likely to be adsorbed in one of 
three different di-$\sigma$-bonded, metastable structures, 
characterized by  
two C-Si bonds, and eventually converts into the 
lowest-energy configurations.
These results are consistent with recent experiments.

\end{abstract}
\vspace{0.5cm}
]

\narrowtext

Adsorption of benzene on the Si(100) surface 
is a topic of great current interest
\cite{Taguchi,Craig,Jeong,Konecny,Gokhale,Borovsky,Self,Lopinski,Kong}
both because it represents a prototype system for the study
of molecular adsorption (and desorption) of hydrocarbons
on semiconductor surfaces, and 
because it is considered a promising precursor for 
technologically relevant processes, such as
the growth of Si-C and CVD diamond thin films on Si surfaces.
However, despite many experimental and theoretical investigations,
the adsorption mechanism is not yet well understood.
In particular, at present there is no consensus 
about the lowest-energy structure of benzene on Si(100):
results obtained from surface science 
experimental techniques, semiempirical methods, and first-principles
approaches provide a number of different predictions.

Benzene is known from experiments to 
adsorb exclusively on top of the Si(100) surface dimer rows, 
thus avoiding energetically disfavored structures 
with unsaturated, isolated Si 
dangling bonds.
Even so, since the size of the benzene molecule is comparable to the spacing
between two adjacent dimers on the same row, many different bonding
configurations are possible. 
Among the structures proposed in the literature 
as the lowest-energy configurations,
the 1,4-cyclohexadiene-like (``butterfly'') configuration,
in which the benzene molecule is di-$\sigma$ bonded to
the two dangling bonds of the same Si surface dimer, 
is supported by thermal desorption and angle-resolved photoelectron
spectroscopy\cite{Gokhale}, STM\cite{Self}, vibrational IR spectroscopy 
and near-edge X-ray absorption fine structure techniques\cite{Kong}, and
first-principles cluster calculations\cite{Gokhale}.
Instead, other STM experiments\cite{Borovsky} suggest the
1,3-cyclohexadiene-like (``tilted'') structure.
Finally, semiempirical calculations\cite{Jeong,Lopinski}, STM and
IR spectroscopy experiments \cite{Lopinski} favor
a tetra-$\sigma$-bonded configuration where benzene is
bonded to two adjacent surface dimers.

Another open issue concerns the occurrence and nature of metastable
adsorption states. In fact, the results of STM and 
IR spectroscopy\cite{Borovsky,Lopinski} 
support the hypothesis that benzene is initially 
chemisorbed in a metastable, ``butterfly''-like state, and then slowly 
converts (within minutes) to a lower-energy final state, which is 
a ``tilted'' structure according to Ref. \onlinecite{Borovsky}, 
or a tetra-$\sigma$-bonded one according to Ref. \onlinecite{Lopinski}.
Moreover, recent IR experiments \cite{Kong} 
suggest that, at room temperature, benzene
is predominantly adsorbed in the ``butterfly'' configuration, 
while the existence of a less stable structure,
consistent with a tetra-$\sigma$-bonded configuration, is proposed.

Previous theoretical calculations on benzene on Si(100) 
have been restricted to semiempirical or {\em ab initio} 
cluster-model methods.   
In the latter approach
the Si surface is modeled with 
a cluster of Si atoms, thus considerably reducing the
cost of a first-principles calculation. However, the effects of such an 
approximation can be relevant. It is well known, for instance, that
the characteristic buckling of the Si dimers on the
clean Si(100) surface can only be obtained by using models with
a slab geometry and periodic boundary conditions.
As shown in the following, the details of the surface reconstruction 
(i.e. buckling and periodicity of the surface dimers) are
crucial ingredients in determining the adsorption structure
of benzene.
Moreover, the convergence
of different properties, such as the binding energies of adsorbed
molecules, is rather slow as a function of
the cluster size.  

In order to overcome these limitations and to clarify the open issues
discussed above we have performed a full 
{\em ab initio} study of benzene adsorption on Si(100).
Total-energy calculations and molecular dynamics 
(MD) simulations have
been carried out within the Car-Parrinello approach\cite{CP,CPMD}
in the framework of the density
functional theory, in the local spin density approximation. 
Tests have been also performed using gradient
corrections in the BLYP implementation\cite{BLYP}.
The calculations have been carried out considering the $\Gamma$-point 
only of the Brillouin zone, and  using norm-conserving 
pseudopotentials\cite{Troullier}, with $s$ and $p$ nonlocality for
C and Si. Wavefunctions were expanded in plane waves with an
energy cutoff of 35 Ry.

The Si(100) surface is modeled with a periodically repeated slab
of 5 Si layers and a vacuum region of 7 \AA~ (tests
have been also carried out with a vacuum region of 10 \AA~, without
any significant change in the results).
A monolayer of hydrogen atoms is used to saturate the dangling bonds on the
lower surface of the slab. 
We have used a supercell with  
$p(\sqrt8\times\sqrt8)R45^{\circ}$ surface periodicity, corresponding to 
8 Si atoms/layer; however, in order to check
finite-size effects, the geometry optimizations have been repeated
using a larger $p(4\times4)$ supercell with 16 atoms/layer.

Structural relaxations of the ionic coordinates are performed 
using the method of direct inversion in the iterative subspace \cite{DIIS}. 
During ionic relaxations and MD simulations
the lowest Si layer and the saturation hydrogens are kept fixed.
We verified that, by starting with the unreconstructed,
clean Si(100) surface, the structural optimization procedure
correctly produces 
asymmetric surface dimers, with a dimer bond length and 
buckling angle in good agreement with previous, highly
converged {\em ab initio} calculations \cite{Bertoni}.
We have considered different surface periodicities for the dimer reconstruction
which may occur on the Si(100)surface, i.e. 
$(2\times 1)$, $p(2\times 2)$ and $c(4\times 2)$.
A single benzene molecule is added on top of the slab and
the system is then fully relaxed towards the minimum energy configuration.
To better explore the complex potential energy surface of this system,
in most of the cases the optimization procedure
was repeated using a simulated-annealing strategy and also starting
from different initial configurations.

We find that the lowest-energy configurations are given by two
tetra-$\sigma$-bonded structures, characterized by the presence
of one C-C double bond, which we refer to as  
``tight bridge'' (TiB) and
``twisted bridge'' (TwB) (see Fig. \ref{struc7}).
TwB is similar to TiB but the benzene molecule is rotated
by 90$^{\circ}$ with respect to the Si surface and is slightly
higher in energy (see Table I).
This result is in agreement with the findings of Ref. \onlinecite{Lopinski}
and turns out to be
independent on the size of the supercell used in the
simulation and on the different reconstructions of the
Si(100) surface. It remains true also using BLYP gradient corrections, 
as can be seen in Table I.

We also find, at somewhat higher energies, three different, 
{\em metastable} ``butterfly'' structures, characterized by two C-Si
bonds, which are shown in Fig. \ref{struc7}.
One of them (``standard butterfly'', SB) is the well-known 
configuration with the
benzene molecule adsorbed on top of a single Si dimer.
The others (``tilted-bridge butterfly'', TB, and ``diagonal-bridge 
butterfly'', DB), which bridge two adjacent surface dimers,
have not been reported in any previous study.

The Si(100) reconstruction crucially affects the occurrence and
energetic ordering of the three ``butterfly'' structures.
In fact, in the $(2\times 1)$ reconstruction (with parallel buckled 
dimers), SB and TB are
the most stable (almost isoenergetic) ``butterfly'' configurations, while
DB is considerably less favored; in contrast, with 
reconstructions involving alternating buckled Si dimers,
such as the $p(2\times 2)$ and the $c(4\times 2)$, SB and DB
are the lowest-energy configurations, while the binding energy of TB is
significantly smaller.
This clearly happens because the two C-Si bonds
of the TB structure are more easily created when the benzene molecule
is adsorbed onto Si(100) $(2\times 1)$, while the formation  
of the DB structure is favored by the presence of alternating buckled 
Si dimers.

The other configurations proposed in the literature, that is 
the ``tilted'' (T) and the ``pedestal'' (P) ones, 
lie higher in energy for all the Si(100) reconstructions considered
(see Table \ref{energy}). In particular,
the P structure is only found to be stable in the
$(2\times 1)$ reconstruction; however, even in this case,
a MD simulation performed at 300 K shows that the 
structure converts very rapidly (in less than 1 ps) into a DB structure.
Although the P structure has four C-Si bonds,
it is energetically disfavored because
it involves the presence of two radical centers.

Inspection of the C-C distances for the various stable structures 
reveals the existence of two kind of bonds: a long one ("single") 
and a short one ("double"), of length $1.49-1.59$ and
$1.34-1.36$ \AA, respectively. These values should be compared with
the C-C bond length in the isolated benzene molecule, 
1.39 \AA.
One double bond characterizes the TiB and TwB structures,
while two double bonds are found in the ``butterfly'' structures.
In contrast, in the P configuration all the C-C bonds are single ones.
These conclusions are confirmed by a more quantitative 
analysis of the electronic orbitals, which we 
performed by using both the notion of Mayer bond order\cite{Mayer}
and the method of the Localized Wannier functions \cite{Wannier}.
In the three ``butterfly'' configurations (SB, TB, DB),
the bond angles ($119^{\circ}$-$122^{\circ}$) at the C atoms not involved 
in the Si-C bonds are close to that ($120^{\circ}$)
of the isolated benzene molecule, while those ($103^{\circ}$-$113^{\circ}$) 
at the 4-fold coordinated C atom are closer
to the ideal tetrahedron ($109.5^{\circ}$) angle.
This clearly indicates $sp^2$ and $sp^3$ hybridization, respectively.  
After benzene chemisorption, although the Si-Si dimers are 
preserved, the Si dimer buckling angle is almost reduced to zero,
with the exception of the TB and DB structures.
In the lowest-energy TiB structure the
angle between the double bond and the Si(100) surface
is $45^{\circ}$ in good agreement with the experimental
estimate \cite{Kong}, $\sim 43^{\circ}$.

The structural parameters do not change appreciably when
a larger $p(4\times4)$ surface supercell is used.
Use of BLYP gradient corrections makes 
bond lengths about 1-2 \% longer, while binding energies 
are significantly reduced (see Table \ref{energy}). 
Moreover, in the $(2\times 1)$ reconstruction, the P
configuration is no longer stable and, 
among the three ``butterfly'' structures, BLYP favors
SB, while the binding energy of DB is even smaller than that of the
T structure.
Note, however, that TiB and TwB remain the lowest-energy configurations.

According to the results of some experiments and theoretical 
calculations\cite{Lopinski,Kong}, adsorbed benzene predominantly forms
a ``butterfly'' (SB) configuration, while the TiB one (and perhaps TwB)
appears in detectable amounts on relatively long timescales only,
thus indicating the existence of an energy barrier between the two
structures.

In order to identify possible metastable states, occurring in the early
stages of adsorption, we have tried to find, in the simplest way,
the most probable
structure of a benzene molecule impinging on the Si(100) surface.
If we place the molecule at some distance from the surface
we observe that, regardless of the initial position and orientation of
the molecule, after full relaxation the final structure 
is almost invariably one of the three 
``butterfly'' configurations.
This happens because the dimers are tilted, favoring
the formation of the di-bonded ``butterfly'' structures rather than
the tetra-bonded ones.
The specific ``butterfly'' configuration which is actually 
formed depends critically on the type of
reconstruction of the Si surface that is considered,
as already discussed above.
On the contrary, there are only very few initial positions which 
lead to the low-energy TiB and TwB configurations.

We have tried to characterize the energy barrier which must be overcome to
relax from the ``butterfly'' configurations
to the lower-energy TiB and TwB structures.  
To this aim we started with the benzene molecule 
in the SB configuration.
Let C$_d$ be one of the C atoms involved in the Si-C bonds.
Many calculations have been performed in which 
the ionic coordinates of both the molecule
and the substrate were optimized under the constraint 
that the $x,y$ coordinates of the two C$_d$ atoms are held fixed.
A particular pathway, connecting the SB to the DB structure, 
is shown in Fig. 2, where the reaction coordinate is defined as 
the distance between the C$_d$-C$_d$ axis of the initial configuration
and that of the displaced structure.
The pronounced energy minimum corresponds to the occurrence, 
during the transformation, of the lowest-energy Tib structure. Note
however that this is characterized by a very narrow well. 
From Fig. 2 a lower bound of $\sim 0.5$ eV can be inferred for the
energy barrier, to be compared with the experimental 
estimates\cite{Borovsky,Lopinski}, $\sim$ 0.9-1.0 eV.
A similar calculation for the TB$\rightarrow$TwB transition gives
a smaller value of $\sim 0.4$ eV. As a consequence the conversion
from TB to TwB is expected to be somewhat faster
than that from SB to TiB.

A large fraction of experiments on benzene on Si(100)
is based on STM techniques. However, different interpretations of 
similar STM images
led to contradictory conclusions\cite{Borovsky,Self,Lopinski}
about the adsorption sites and geometry of the adsorbed molecules.
For each of the structures reported in Table I we have produced
``theoretical'' STM images to be compared with the experimental ones,
following the recipe of Ref. \onlinecite{Takeuchi}.
Charge density iso-surfaces have been obtained by including 
electron states in an
energy range down to $\sim 2$ eV below the highest 
occupied state, which
corresponds to typical STM bias voltages. The simulated
images are obtained by viewing these iso-surfaces at typical 
tip-surface distances (a few \AA\ above the benzene molecule).

Our computed STM image for the TiB structure exhibits a density
maximum above one of the two Si dimers involved in bonding with
benzene, while the TwB configuration produces a similar
image but rotated by 90$^{\circ}$.
These images resemble those obtained by Lopinski 
{\em et al.}\cite{Lopinski}.
The theoretical STM image for the SB structure is characterized
by a bright two-lobe protrusion centered symmetrically above 
a single Si dimer unit and
oriented orthogonal to the dimer axis, in qualitative agreement with
the experimental findings\cite{Borovsky,Self,Lopinski}.
Instead, the STM images of the TB and DB structures are quite different
from that of SB. In fact the TB image is qualitatively similar to 
that of TwB
(and the experimental STM resolution could be insufficient to 
distinguish between the two configurations), while
DB gives rise to a much fainter feature, 
bridging in diagonal two Si dimers,
which is probably hardly visible in experiments.
These observations could explain why the DB and TB structures
have not been detected in STM experiments.
The T configuration produces an asymmetric
(with respect to Si dimers) image, appearing as 
a bright region (placed between two Si dimers) adjacent to a dark region. 
Finally the P structure is characterized 
by two spots corresponding to the 
dangling bonds of benzene; this result supports the conjecture
\cite{Self} which rules out the presence of a
significant fraction of benzene molecules adsorbed in the 
P structure because of the absence of such spots in the STM images.

We have also computed the vibrational spectra for a representative 
``butterfly'' structure, SB, and for the lowest-energy TiB configuration, 
by performing Car-Parrinello MD simulations at room temperature. 
Our results for TiB show a slightly more 
quantitative agreement with the experimental results
\cite{Taguchi,Kong} than those for the SB structure, although 
the main features of the spectra are similar in the two structures.
Let C$\,''$ (C$\,'$) denote a C atom which shares a double
(single) bond with another C atom.
The C$\,'$-H and C$\,''$-H frequencies (2880 and 3010 cm$^{-1}$) are  
in agreement with
the $sp^3$ and $sp^2$ stretching modes observed
in recent IR spectroscopy experiments \cite{Kong}
(2945 and 3044 cm$^{-1}$), and semiempirical
cluster calculations \cite{Lopinski}.
Note that the C-H vibrations for the isolated benzene molecule are 
characterized by a single frequency of 3040 cm$^{-1}$.
For the C$\,'$-C$\,''$ and C$\,''$-C$\,''$ frequencies we find 
1230 and 1520 cm$^{-1}$, respectively, to be compared with 
the EELS experimental values \cite{Taguchi}, 1170 and 1625 cm$^{-1}$.
The C-H bending modes are found at 900 and 1100 cm$^{-1}$,
whereas experimentally \cite{Taguchi} they are at 910 and 1075 cm$^{-1}$. 

In conclusion, using state-of-the-art {\em ab initio} simulations,
we have shown that a tetra-$\sigma$ bonded structure 
is the most stable configuration for benzene adsorbed on Si(100).
However, this structure and a very similar one, lying only slightly 
higher in energy, correspond
to very narrow wells in the potential energy surface for
a benzene molecule impinging on the surface. Therefore
it is more likely for the molecule to be adsorbed
into one of three different, metastable ``butterfly'' configurations,
and eventually convert into the lowest-energy structures.
Our study provides detailed information about structural,
electronic, and vibrational properties of the system,
and allows a critical comparison with results obtained 
from different experimental techniques and previous theoretical
calculations. 

We thank M. Boero and A. Vittadini for useful discussions.
This work is partially supported by INFM
through the Parallel Computing Initiative.

\vfill
\eject

\begin{figure}
\caption{The stable structures of benzene adsorbed on Si(100):
SB=``standard butterfly'', TB=``tilted-bridge butterfly'',
DB=``diagonal-bridge butterfly'', T=``tilted'', P=``pedestal'',
TiB=``tight bridge'', TwB=``twisted bridge''.
For clarity only the four Si atoms of two dimers and
four belonging to the second layer are shown.} 
\label{struc7}
\end{figure}

\begin{figure}
\caption{Total energy along the pathway obtained by shifting the benzene
molecule along a dimer row from the SB (at the origin)
to the DB configuration, going through the lowest-energy TiB configuration
(on the bottom of the narrow well). 
A $p(\sqrt8\times\sqrt8)R45^{\circ}$ supercell with a $p(2\times 2)$
surface reconstruction has been used.
Data are represented by symbols, while the line
is just a guide for the eye.
The energies are relative to the SB structure.}
\label{path}
\end{figure}

\vfill
\eject

\begin{table}
\caption{Binding energies (in eV) of different configurations for
benzene adsorbed on Si(100) in the $(2\times 1)$ and $c(4\times 2)$
reconstructions (the nomenclature is the same as in Fig. 1).
The $p(\sqrt8\times\sqrt8)R45^{\circ}$ supercell was used; $L$ denotes
results obtained with the larger $p(4\times4)$ supercell and BLYP
means application of BLYP gradient corrections [12]. 
A missing entry
indicates that a stable configuration was not obtained by the optimization
process.}
\begin{tabular}{lllll}
Configuration&$(2\times 1)$&$(2\times 1)$ $L$&$(2\times 1)$ BLYP&$c(4\times 2)$ $L$\\ \tableline
SB&2.04&2.06&1.22&2.20 \\
TB&2.10&2.08&1.12&1.99 \\
DB&1.63&1.70&0.41&2.24 \\
T& 1.50&1.55&0.77&1.68 \\
P& 1.51&1.60&---&--- \\
TiB& 2.68&2.77&1.53&2.65 \\
TwB& 2.47&2.53&1.31&2.38 \\
\end{tabular}
\label{energy}
\end{table}

\vfill
\eject

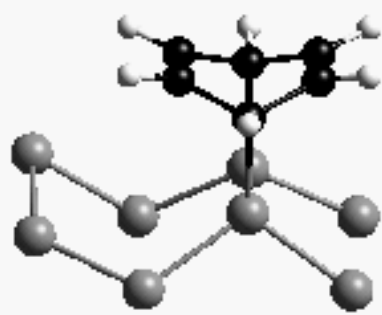
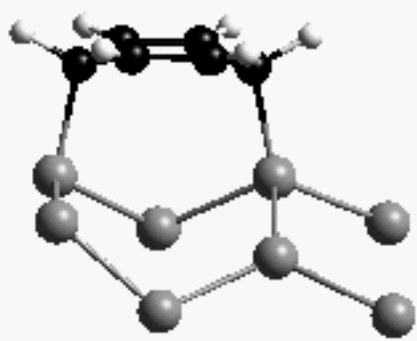
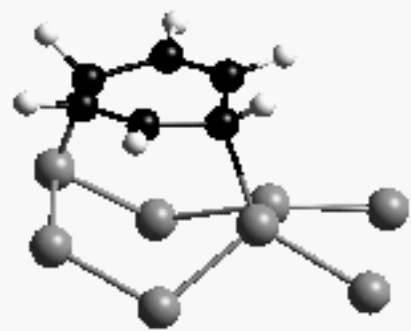
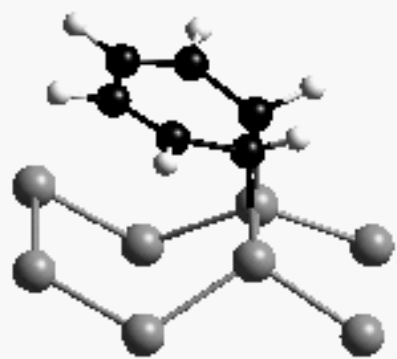
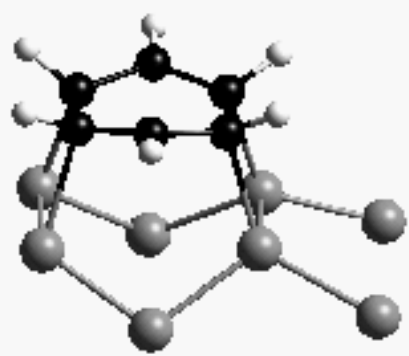
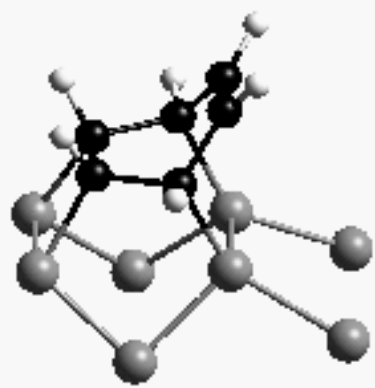
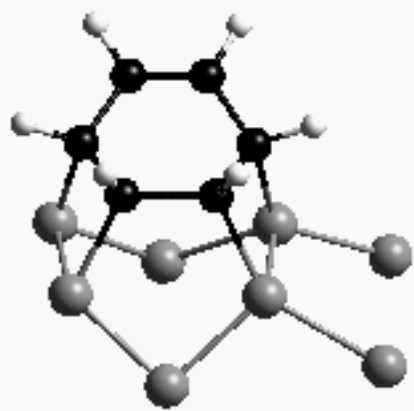
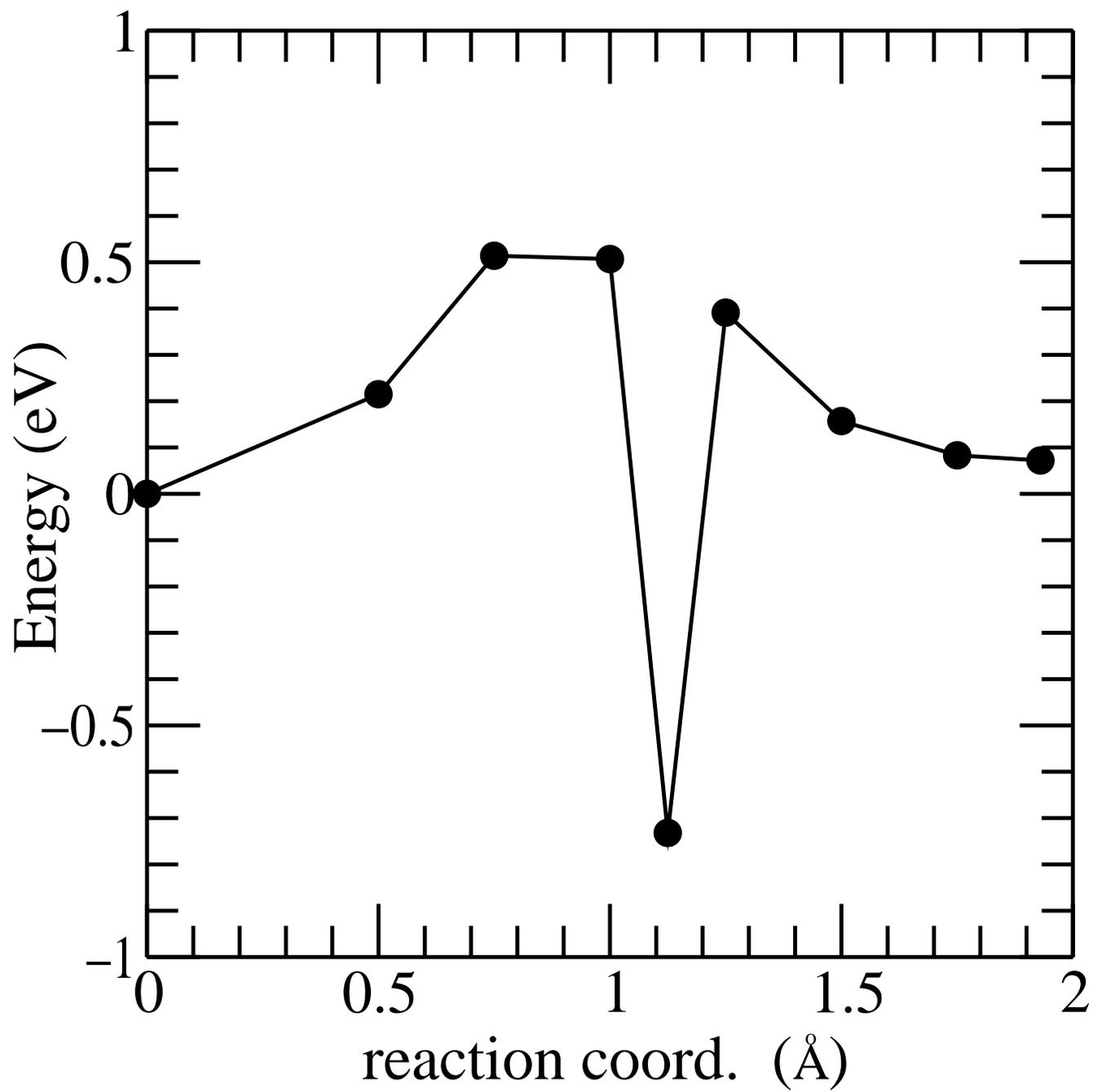
\end{document}